# Nuclear spin temperature reversal via continuous radio-frequency driving

Pablo R. Zangara[1,2,*], Daniela Pagliero[3,*], Ashok Ajoy[5,7], Rodolfo H. Acosta[1,2], Jeffrey A. Reimer[6,7], and Carlos A. Meriles[3,4,†]

[1]*Universidad Nacional de Córdoba, Facultad de Matemática, Astronomía, Física y Computación, Córdoba, Argentina.*
[2] *CONICET, Instituto de Física Enrique Gaviola (IFEG), Córdoba, Argentina.*
[3]*Department. of Physics, CUNY-City College of New York, New York, NY 10031, USA.*
[4]*CUNY-Graduate Center, New York, NY 10016, USA.*
[5]*Department of Chemistry, University of California at Berkeley, Berkeley, California 94720, USA.*
[6]*Department of Chemical and Biomolecular Engineering, University of California at Berkeley, Berkeley, California 94720, USA.*
[7]*Materials Science Division Lawrence Berkeley National Laboratory, Berkeley, California 94720, USA.*

[*]*Equally contributing authors*

[†]*Corresponding author. E-mail: cmeriles@ccny.cuny.edu.*

Optical spin pumping of color centers in diamond is presently attracting broad interest as a platform for dynamic nuclear polarization at room temperature, but the mechanisms involved in the generation and transport of polarization within the host crystal are still partly understood. Here we investigate the impact of continuous radio-frequency (RF) excitation on the generation of nuclear magnetization produced by optical illumination. In the presence of RF excitation far removed from the nuclear Larmor frequency, we witness a magnetic-field-dependent sign reversal of the measured nuclear spin signal when the drive is sufficiently strong, a counter-intuitive finding that immediately points to non-trivial spin dynamics. With the help of analytical and numerical modeling, we show our observations indicate a modified form of 'solid effect', down-converted from the microwave to the radio-frequency range through the driving of hybrid transitions involving one (or more) nuclei and two (or more) electron spins. Our results open intriguing opportunities for the manipulation of many-electron spin systems by exploiting hyperfine couplings as a means to access otherwise forbidden intra-band transitions.

## I. INTRODUCTION

Although the principles underlying dynamic nuclear polarization (DNP) — in its many incarnations — were introduced more than half a century ago[1], the last two decades have witnessed a sustained, widespread interest in the topic[2-7]. This trend can largely be seen as a response to fundamental, well-known traits of nuclear magnetic resonance (NMR), outstanding in its versatility and range of applications but inherently limited in its detection sensitivity. Understandably, much of the effort has been devoted to developing dynamic polarization strategies applicable to high magnetic fields, a route that promises to reap the benefits of enhanced NMR signals without sacrificing on spectral resolution[8-10]. On the other hand, low-field setups — especially those based on optical pumping and detection — allow the experimenter to study regimes otherwise difficult to access[11,12]. Particularly important is the ability to probe the dynamics of nuclear spins adjacent to paramagnetic defects, typically invisible in traditional high-field DNP experiments[13-15].

From among various existing routes, optical spin pumping of paramagnetic defects in wide-bandgap semiconductors has garnered recent attention as a broadly applicable DNP platform, largely due to the potential for room temperature (or near room temperature) operation[16-20]. Often hosting a dilute set of nuclear spins, materials such as diamond or silicon carbide have also emerged as versatile systems to investigate fundamental questions on the spin dynamics governing DNP[21-24], most notably the transport of spin polarization from strongly-hyperfine-coupled to bulk nuclei. For example, diamond was recently exploited to show how coupled electron spin networks can effectively mediate interactions between remote nuclei, in the process circumventing the transport restrictions created by the 'spin diffusion barrier' surrounding individual paramagnetic defects[25-28].

Here we extend prior work on optical spin pumping of NV-hosting diamond to investigate the impact of a continuous radio-frequency (RF) drive on the generation of $^{13}$C spin polarization via electron spin cross relaxation. We focus on strongly hyperfine-coupled carbons and experimentally show that sufficiently strong RF excitation during optical spin pumping can invert the sign of the observed nuclear polarization. With the aid of numerical modeling, we interpret this counter-intuitive finding as a consequence of electron/nuclear spin mixing, allowing us to RF-drive typically hindered transitions between electron spin states with (nearly) the same Zeeman energy. Examining the system response at variable magnetic fields, we find this process emerges from a subtle interplay between the type and relative concentrations of paramagnetic defects present in the sample, the number of nuclei featuring a given hyperfine coupling, and their spin transport efficiency at the applied magnetic field.

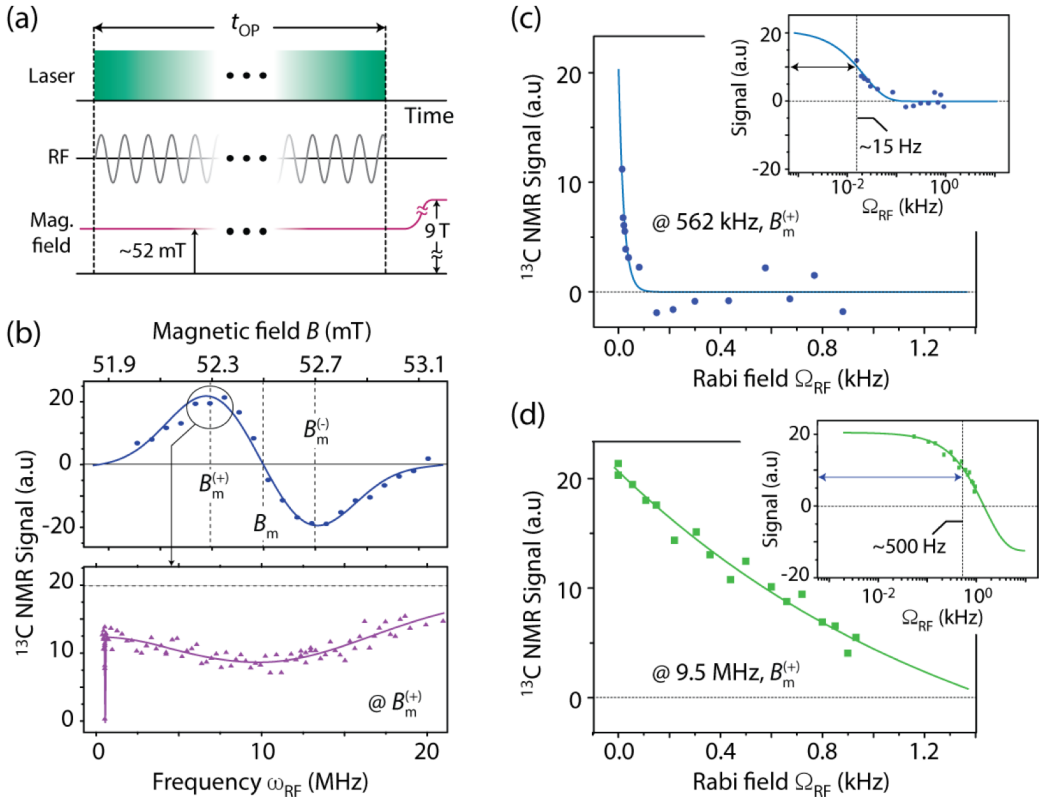

**Fig. 1: Dynamic nuclear polarization in the presence of an RF drive.** (a) We use a field cycling protocol to monitor the $^{13}$C polarization induced by optical spin pumping at a low magnetic field $B$ in the presence of an RF drive of variable frequency. (b) (Top) $^{13}$C NMR signal amplitude as a function of $B$ without RF. NV-P1 cross polarization leads to positive (negative) polarization below (above) the matching field $B_m = 52.5$ mT. (Bottom) $^{13}$C signal amplitude at $B_m^{(+)}$ as a function of RF frequency for a constant (i.e., frequency-insensitive) RF drive of amplitude $\Omega_{RF} = 0.7$ kHz. In both panels, solid lines are guides to the eye. (c) $^{13}$C signal amplitude as a function of RF power for a drive frequency ($\omega_{RF} = 560$ kHz) resonant with the $^{13}$C Larmor frequency at $B_m^{(+)}$. The signal decreases to 50% the original amplitude with an RF amplitude as low as $\Omega_{RF} = 15$ Hz (insert). The solid line indicates an exponential fit. (d) Same as in (c) but for an RF frequency of 9.5 MHz. As inferred from the exponential fit (solid line), we observe a slower decrease with RF power and a trend toward negative signal amplitudes.

## II. RESULTS

We work under conditions similar to those reported previously[26,31]. Briefly, we implement a field-cycling sequence that starts with simultaneous laser illumination and RF drive of variable frequency and amplitude for a time interval $t_{OP}$, followed by sample shuttling to a high-field magnet (Fig. 1a). We probe the $^{13}$C magnetization using one-pulse excitation at the $^{13}$C Larmor frequency at 9 T (~100 MHz) to produce a free induction decay (FID); upon Fourier transform, we monitor the peak amplitude of the in-phase component (referred to below as the '$^{13}$C NMR signal'), adjusted to yield a positive (negative) sign when the dynamically-pumped and thermal $^{13}$C magnetizations are parallel (anti-parallel). In all instances, we choose $t_{OP} = 5$ s — i.e., comparable to the $^{13}$C spin-lattice relaxation at the pumping field — so as to attain maximum signal amplitude.

For the present experiments, we use a high-pressure/high-temperature (HPHT) [100] diamond crystal engineered to host nitrogen-vacancy (NV) centers at a concentration of ~10 ppm; green illumination (532 nm, 1 W) spin-pumps NV centers into the $m_S = 0$ state of its ground spin triplet ($S = 1$), which we ultimately exploit to dynamically polarize $^{13}$C spins. To this end, we set the magnetic field during $t_{OP}$ somewhere near the matching field $B_m$ — 52.5 mT under our present experimental conditions — where the energy separation between the $|m_S = 0\rangle$ and $|m_S = -1\rangle$ states of the NV coincides with the Zeeman splitting of coexisting spin-1/2 paramagnetic defects. Particularly relevant herein are the so-called P1 centers — point defects formed by neutral substitutional nitrogen featuring a spin number $S' = 1/2$ — present in our sample at a concentration of 50 ppm. Slightly below (or above) $B_m$, a simultaneous NV–P1 'double flip' — corresponding to a transition from $|m_S = 0, m_{S'} = +1/2\rangle$ to $|m_S = -1, m_{S'} = -1/2\rangle$ — polarizes adjacent $^{13}$C spins positively (or negatively), as dictated by spin energy conservation (top panel in Fig. 1b). Repeated NV initialization followed by three-spin flips and nuclear spin diffusion to bulk (i.e., non-hyperfine-coupled) carbons ultimately leads to the observed sample magnetization. We refer the reader to Refs. [29–32] for a more thorough discussion on $^{13}$C-enabled NV–P1 cross-relaxation.

While the physical picture above provides an intuitive starting point, the mechanisms allowing nuclear polarization to spread to bulk nuclei are far from simple. Part of the problem is the large energy mismatch between nuclei featuring different hyperfine couplings, a regime that truncates nuclear flip-flop terms in the Hamiltonian and thus (presumably) quenches spin diffusion. We found in prior work[26] that, contrary to expectations, nuclear spins strongly coupled to the NV contribute to transporting polarization to the bulk, a process mediated by interactions between electronic spins. For completeness, we reproduce part of these results in Fig. 1b (lower panel), where we plot the $^{13}$C NMR signal at $B_m^{(+)}$, here defined as the magnetic field where the carbon polarization is maximum, see upper panel in Fig. 1b. Under an RF drive of fixed amplitude (corresponding to a $^{13}$C Rabi frequency $\Omega_{RF} = 0.7$ kHz) and variable frequency $\omega_{RF}$, we find a broad RF absorption spectrum that can be related to the family of hyperfine-coupled carbons[26] (found in second- and higher-order atomic shells around NVs[33-35]).

Although the above observations expose the important role strongly-hyperfine-coupled carbons play in transporting nuclear polarization, the exact nature of the RF absorption at play — meaning the type of spin levels we couple via the drive — remains unclear. In the most intuitive picture, RF excitation drives nuclear-spin-only transitions (i.e., transitions where the electron spin projections are conserved), thus causing 'saturation' (i.e., effective equilibration between nuclear spin population differences) and, correspondingly, a reduction of the observed $^{13}$C NMR signal. This process is most clearly exposed at ~560 kHz — associated to the Larmor frequency of bulk nuclei — where weak cw excitation during $t_{OP}$ is sufficient to extinguish the observed nuclear polarization (Fig. 1c). An immediate question therefore arises, namely, is this same description applicable to hyperfine-coupled carbons as well? Initial observations at higher frequencies suggest this is not the case. An illustration is presented in Fig. 1d where we plot the $^{13}$C NMR signal for an RF drive at 9.5 MHz. Relative to the results in Fig. 1c, we find a much slower decrease indicating the RF drive is not as efficient in saturating the set of spin transitions resonant at this frequency. Perhaps more intriguingly, an extrapolation to higher RF powers — beyond our reach in this early set of experiments — suggests that the sign of the NMR signal reverses for sufficiently strong drives.

To further investigate these observations, we altered the probe in our shuttling experiments — initially designed to yield RF excitation of uniform, nearly-frequency-independent amplitude over a range reaching 150 MHz[26] — to feature a resonance around $\omega_{RF}^{Res} \approx 11.5$ MHz, approximately coincident with the range of transition frequencies expected for second shell carbons[33-35]. Even if at the expense of a non-uniform response, this setup allows us to generate RF amplitudes (near $\omega_{RF}^{Res}$) beyond those possible in our prior design, thus allowing to probe strongly-hyperfine-coupled nuclei in the unexplored regime of strong drives.

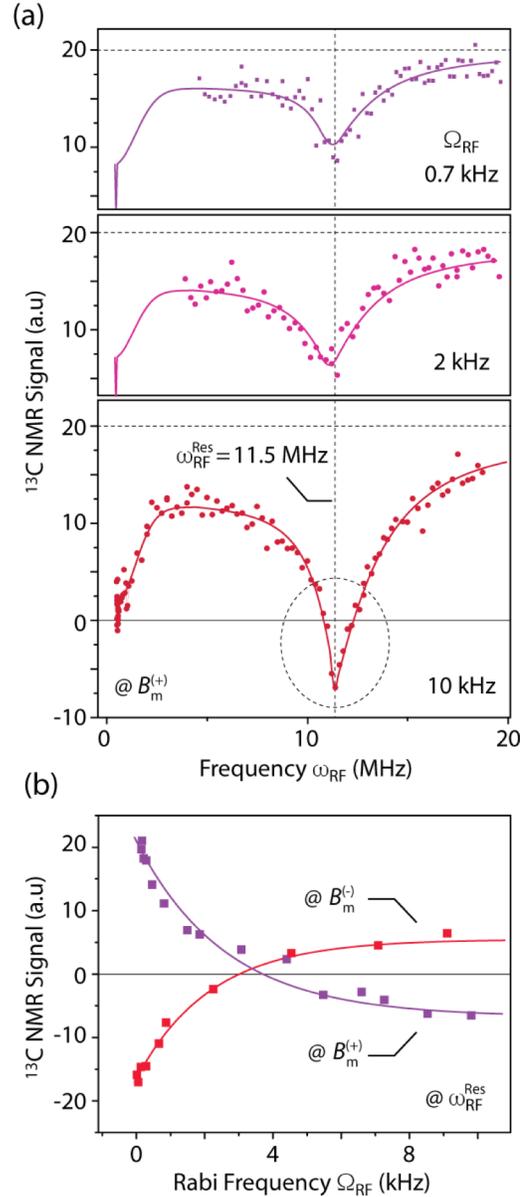

**Fig. 2: The limit of strong RF drive.** (a) $^{13}$C signal amplitude at $B_m^{(+)}$ as a function of RF frequency for a modified RF probe featuring a resonance at 11.5 MHz. At the highest RF power ($\Omega_{RF} = 10$ kHz at 11.5 MHz), we observe inversion of the NMR signal. (b) $^{13}$C signal at $B_m^{(+)}$ and $B_m^{(-)}$ (purple and red squares, respectively) as a function of the RF power for a drive at 11.5 MHz.

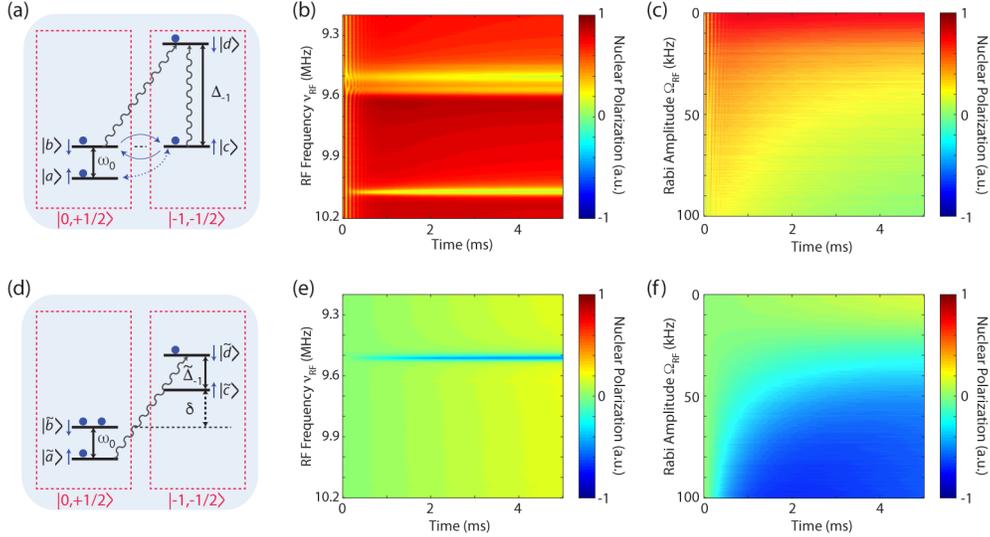

**Fig. 3. Dynamic nuclear polarization in the presence of an RF drive.** (a) We consider a $^{13}$C–NV–P1 set where a $^{13}$C nucleus is hyperfine-coupled to an NV, in turn, experiencing a dipolar interaction $\mathcal{J}_d$ with a neighboring P1. We restrict the energy diagram to eigenstates $\{|a\rangle, |b\rangle, |c\rangle, |d\rangle\}$ within the manifold $|m_S = 0, m_{S'} = +1/2\rangle$ and $|m_S = -1, m_{S'} = -1/2\rangle$ and use blue $\{\uparrow, \downarrow\}$ arrows to denote nuclear spin projections; $\omega_0$ is the $^{13}$C Larmor frequency, and $\Delta_{-1} = \sqrt{(A_{zx})^2 + (A_{zz} + \omega_0)^2}$ where $A_{zz}$ and $A_{zx}$ denote the secular and pseudo-secular hyperfine coupling constants. We set the magnetic field to $B_m^{(+)}$, where states $|b\rangle$ and $|c\rangle$ are degenerate and assume that optical pumping initializes the system in the $|0, +1/2\rangle$ manifold; under strong RF (wavy arrows), the otherwise positive nuclear polarization developing at $B_m^{(+)}$ (blue arcs) is destroyed without, however, producing polarization inversion. Note the weak exchange between states $|a\rangle$ and $|c\rangle$, responsible for the resulting zero net magnetization (see text). (b) Calculated $^{13}$C polarization as a function of time and RF frequency assuming $A_{zz} = A_{zx} = 6.5$ MHz and $\mathcal{J}_d = 100$ kHz; the RF Rabi amplitude is $\Omega_{RF} = 50$ kHz. (c) Calculated $^{13}$C polarization as a function of time and $\Omega_{RF}$ for a cw drive at $\omega_{RF} \sim 9.5$ MHz. (d) Same as in (a) at the same field $B_m^{(+)}$ but for a $^{13}$C–NV–P1 set featuring a hyperfine splitting $\tilde{\Delta}_{-1}$; we use tildes throughout the diagram as a reminder this is not the spin system considered in (a). (e) Same as in (b) but assuming $\tilde{A}_{zz} = \tilde{A}_{zx} = 5.7$ MHz, which yields a mismatch $\delta \approx \omega_0$ between states $|\tilde{b}\rangle, |\tilde{c}\rangle$. (f) Same as in (c) but assuming the system in (d) and a cw drive at $\omega_{RF} \sim 9.5$ MHz. Strong RF excitation induces polarization inversion.

Fig. 2a reproduces the measured RF absorption spectrum observed with this modified setup. We find an altered frequency dependence with a clear dip centered at $\omega_{RF}^{Res}$, a consequence of the locally-enhanced RF transmission. Remarkably, we witness — to our knowledge, for the first time — a sign reversal in the crystal polarization when the RF amplitude is sufficiently strong (lower trace in Fig. 2a). As shown in Fig. 2b, the signal inversion is partial and saturates to about one fourth of the starting amplitude when the RF Rabi field reaches 10 kHz. Further, we find that changing the magnetic field from $B_m^{(+)}$ to $B_m^{(-)}$ — see notation in the upper panel of Fig. 1b — produces a mirrored response, where the system polarization (negative in the absence of RF) gradually reaches a partially inverted amplitude (red trace in Fig. 2b).

While simple in its exterior, the finding above is profound in its implications. First off, we note that the continuous wave (cw) nature of the RF excitation (in all experiments, $t_{OP} = 5$ s) strongly suggests that the observed signal inversion cannot derive from coherent spin manipulation: As shown in Appendix A, such a scenario would require a near-perfectly timed process of polarization generation, storage, and transport, an unlikely mechanism given the diffusive nature of the dynamics in all steps. On the contrary, we hypothesize simultaneous RF and NV optical pumping drive hyperfine-coupled nuclei into a steady state of inverted spin temperature, ultimately spreading to bulk carbons and correspondingly reversing the sign of the observed magnetization.

To more clearly lay out our theoretical framework, we return to the $^{13}$C–NV–P1 system considered above and focus our attention on the set of (nearly degenerate) states $\{|a\rangle, |b\rangle, |c\rangle, |d\rangle\}$ within the $|m_S = 0, m_{S'} = +1/2\rangle$ and $|m_S = -1, m_{S'} = -1/2\rangle$ manifolds (see Fig. 3a and Appendix B). For clarity, we first consider the case where states $|b\rangle$ and $|c\rangle$ share the same energy. Provided optical pumping initializes the system into equally populated states $|a\rangle$ and $|b\rangle$, positive nuclear spin polarization follows from the population exchange between states $|b\rangle$ and $|c\rangle$ (accompanied by spin diffusion[31]). With the understanding that the magnetic field enabling this process depends on the assumed magnitude of the hyperfine coupling, in our simulations we generically associate the above spin dynamics to $B_m^{(+)}$.

To quantitatively model the system evolution in the simultaneous presence of RF and optical excitation, we transform the $^{13}$C–NV–P1 Hamiltonian to an effective rotating frame, and implement a quantum jump Monte Carlo that stochastically projects the NV into $|m_S = 0\rangle$ (Appendix B). Fig. 3b shows the $^{13}$C spin polarization as a function of time for an RF drive of variable frequency assuming the spin set starts from an equal mixture of states $|a\rangle$ and $|b\rangle$. The system quickly evolves into a steady state of positive nuclear polarization except at select frequencies (or frequency bands) where the RF drive leads to near-complete saturation. In each case, the nature of the transition at play can be (qualitatively) established with the help of the system's energy diagram. An illustration is presented in Fig. 3c where we analyze the polarization dip near 9.5 MHz. In the spin set assumed for these simulations, this frequency corresponds to the hyperfine splitting $\Delta_{-1}$ between states $|c\rangle$ and $|d\rangle$ (see Appendix B), implying that a sufficiently strong RF drive can quickly redistribute spin populations and hence preempt the generation of nuclear polarization. Note that because states $|b\rangle$ and $|c\rangle$ are degenerate at the chosen magnetic field, RF excitation also induces a population exchange between states $|b\rangle$ and $|d\rangle$, belonging to different electronic spin manifolds. Further, the proximity between states $|a\rangle$ and $|c\rangle$ — separated only by $\omega_0$ — also leads to a weak population exchange (dotted line in Fig. 3a). The result is a complete saturation of the nuclear spin polarization (lower right corner in Fig. 3c) without leading, however, to signal inversion.

From the above considerations, we conclude the steady state nuclear polarization emerges from an intricate interplay between the NV–P1 dipolar coupling (governing the exchange rate between $|b\rangle$ and $|c\rangle$) and the RF amplitude (connecting $|d\rangle$ with states $|c\rangle$ and $|b\rangle$). The calculated time evolution for variable RF amplitude $\Omega_{\rm RF}$ is shown in Fig. 3c assuming an NV–P1 coupling $\mathcal{J}_d = 100$ kHz. The system reaches a steady state after only a few milliseconds though the polarization remains positive at all times, even under strong RF[*]. It is worth mentioning that a similar physical picture — leading to polarization saturation without signal inversion, not included here for brevity — applies to the dip at ~10.1 MHz in Fig. 3b, this time associated to an RF-induced transition between states $|a\rangle$ and $|d\rangle$.

To address the question as to how an RF drive induces negative nuclear polarization, we consider the dynamics of a $^{13}$C–NV–P1 spin set with a different $^{13}$C–NV coupling — i.e., featuring a different hyperfine splitting $\widetilde{\Delta}_{-1}$ — at the same applied magnetic field (Fig. 3d). For future reference, here we choose the hyperfine coupling so that $\delta$ — the energy mismatch between states $|\tilde{b}\rangle$ and $|\tilde{c}\rangle$ — amounts to $\omega_0$, though this condition is, in general, unnecessary. The calculated $^{13}$C polarization as a function of time at a variable RF frequency is shown in Fig. 3e: In the presence of a mismatch, virtually no net magnetization develops over time except near 9.5 MHz, where the $^{13}$C spin polarizes negatively.

The schematic in the diagram of Fig. 3d reveals the dynamics at play, namely, nuclear polarization emerges from a selective population exchange between $|\tilde{a}\rangle$ and $|\tilde{d}\rangle$. Since the latter correspond to states where all spins (both nuclear and electronic) have different projections, this transition (otherwise prohibited) is enabled here thanks to the hybridization caused by hyperfine and electron-electron interactions (see Appendix B). Correspondingly, the RF amplitude required to induce full polarization inversion — in general dependent on the specifics of the modelled spin set — must grow as the inter-particle couplings become weaker. We note that contrary to what the simplified energy diagram in Fig. 3a may suggest, similar considerations apply to all dips in Fig. 3b as well because the degeneracy between $|b\rangle$ and $|c\rangle$ creates a situation where nuclear spin projections (maintained in the diagram for simplicity) lose their meaning (both combine $|\uparrow\rangle$ and $|\downarrow\rangle$ nuclear spin states as well as $|0, +1/2\rangle$ and $|-1, -1/2\rangle$ contributions). The consequence is that the RF amplitude required to saturate the $^{13}$C polarization is analogous to that needed to reach the inverted steady state, as a comparison between Figs. 3c and 3f confirms.

Out of the two spin sets we consider, only one of them polarizes positively when no RF is present (Figs. 3a-3c), while the other one acts as a source of negative polarization just when the RF is on (Figs. 3d-3f). Both exhibit dips at the same frequency (~9.5 MHz in this example, a consequence of having chosen $\delta = \omega_0$), meaning that for the observed signal to reverse sign, the RF drive must gradually shift the leading role away from the 'matched' set (saturating as the RF increases, Fig. 3c) to the group of 'mismatched' spins (evolving from passive bystanders to the source of inverted polarization, Fig. 3g). Therefore, in an experiment where the magnetic field and RF drive are constant — as in the present case — the sign of the observed bulk polarization emerges from a statistical average between contributions from both types of spin sets.

It is interesting to compare the above process with the well-known "solid-effect"[36]: In the simplest case where the electronic and nuclear spin numbers are both equal to 1/2, nuclear polarization emerges as microwave excitation drives "forbidden" transitions between states whose electron and nuclear spin projections are different. Something similar can be said about the spin subspace in the $^{13}$C–NV–P1 set of Fig. 3d except that the near-degeneracy between the $|0, +1/2\rangle$ and $|-1, -1/2\rangle$ states allows us to bring down the excitation frequency to the RF range. As discussed above, this double electron spin flip is allowed via the NV–P1 dipolar coupling so the proposed mechanism simultaneously capitalizes on the "solid" and "cross" effects.

Fig. 4a shows an extension of the experiments in Fig. 2 to variable magnetic fields and RF drive of fixed frequency ($\omega_{\rm RF}^{\rm Res} = 11.5$ MHz). Consistent with the observations in Fig.

---

[*] Mild signal inversion was seen in these simulations as a result of the Bloch-Siegert effect, but the required Rabi fields — in excess of 200 kHz — are beyond those attained in our present experiment and cannot be invoked to explain our observations.

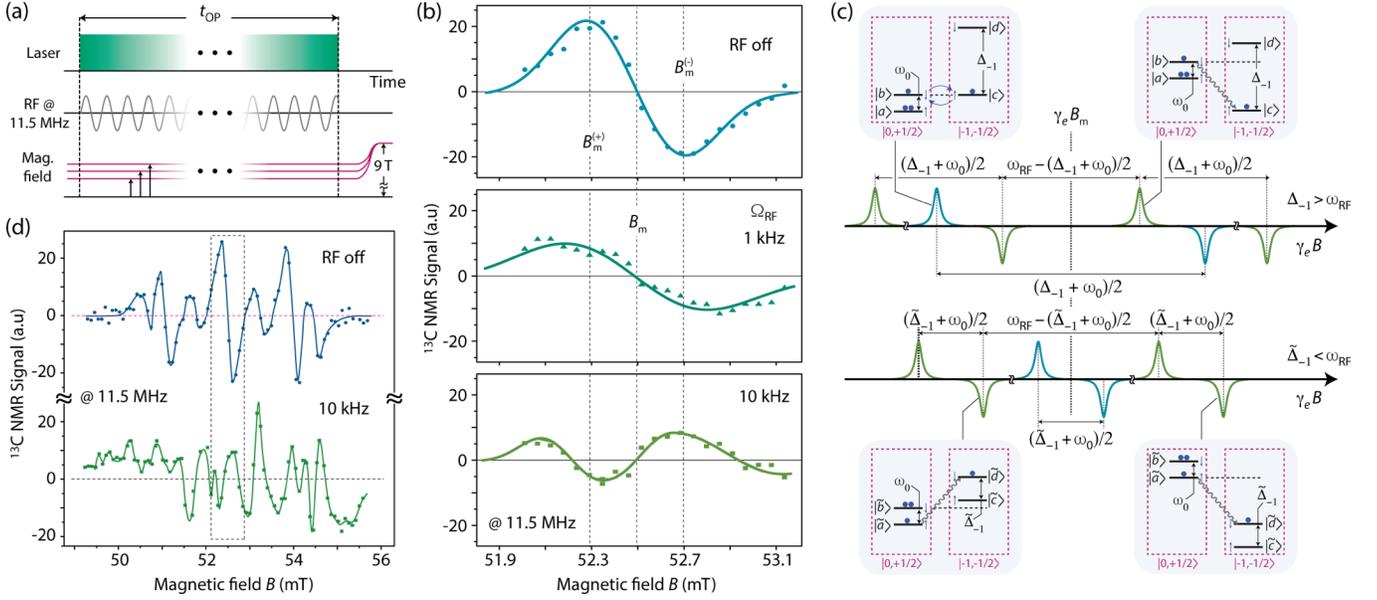

**Fig. 4. The interplay between RF drive and applied magnetic field.** (a) Experimental protocol. We monitor the $^{13}$C DNP signal under a continuous RF drive at $\omega_{\text{RF}}^{\text{Res}} = 2\pi \times 11.5$ MHz of variable amplitude for different applied magnetic fields. (b) $^{13}$C NMR signal as a function of $B$ near $B_m$ for variable RF amplitude. Signal reversal is observed both near $B_m^{(+)}$ and $B_m^{(-)}$, though the resulting pattern lacks overall symmetry. Solid lines are guides to the eye. (c) Schematics of $^{13}$C polarization for the three-spin model as a function of the electron Larmor frequency $\gamma_e B$ assuming an RF drive of frequency $\omega_{\text{RF}}$. We use light green (blue) traces to indicate polarization stemming from the RF-drive (NV–P1 cross-relaxation). The upper (lower) diagram refers to a $^{13}$C–NV–P1 spin set where the hyperfine splitting is greater (smaller) than $\omega_{\text{RF}}^{\text{Res}}$; as before, we use tildes to differentiate eigenstates from one set or the other. Energy or frequency separations are not to scale. (d) Same as in (b) but for a wider magnetic field range that includes all $^{14}$N-induced hyperfine shifts. The Rabi field is 10 kHz in the lower trace and negligible in the upper trace.

2b, we find that the $^{13}$C signal reverses sign near $B_m^{(\pm)}$, though a comparison of the full response shows a non-uniform change marked by the gradual disappearance of signal inversion for large magnetic field shifts relative to $B_m$ (Fig. 4b). Qualitatively, we relate this dependence to the combined effect of contributions from heterogeneous spin sets, each reacting differently to the RF drive and applied magnetic field. We illustrate this interplay in Fig. 4c where we consider the nuclear polarization in $^{13}$C–NV–P1 spin sets featuring different hyperfine couplings: Besides the expected contributions from energy matching (represented as blue positive and negative Lorentzians), RF excitation leads to additional 'peaks and dips' whose relative position depends on the ratio between the hyperfine splitting $\Delta_{-1}$ and the RF frequency $\omega_{\text{RF}}$ (green Lorentzians). Sufficiently far away from $B_m$, contributions stemming from the RF drive have the same sign as those derived from field matching implying that no reversal of the observed bulk signal is expected, in agreement with experiment. Closer to $B_m$, however, overlaps between positive and negative contributions from different spin sets are unavoidable. The result is that the sign of the observed signal can only derive from a subtle interplay that factors in not only the statistical abundance of a given hyperfine coupling but also the RF efficiency in driving a forbidden transition as well as the transport dynamics from the relevant spin set into bulk nuclei.

We emphasize that the above physical picture must be seen as a crude simplification, as the three-spin system studied herein serves only as a footprint for more complex interacting sets — likely involving multiple electron and nuclear spins — whose formation largely depends on the sample composition. A strong indication supporting this view can be found in the pattern of Fig. 4b, twice as broad compared to similar observations in other diamond samples[32]. While larger arrays will likely exhibit analogous spin dynamics, a quantitative connection between the observed RF absorption dips and the exact type of the transitions involved becomes increasingly difficult — if not impossible — without a detailed sample characterization. This problem becomes all the more apparent in Fig. 4c, where we extend the sampled magnetic field range to encompass the full set of NV–P1 transitions (each centered around a 'matching field' whose exact value depends on the nuclear spin projection of the host $^{14}$N spin in each defect[31,32]). While signal inversion clearly extends beyond the central range, the irregular response we find fully exposes the complexity of the problem at hand. Future work — likely relying on diamond crystals with a lower nitrogen content where simpler spin sets become

dominant — will therefore be needed to gain a more quantitative understanding.

## III. CONCLUSIONS

While the basic ideas governing dynamic nuclear polarization have long been established, a broad effort is ongoing to shed light on the microscopic mechanisms involving the generation and transport of polarization to bulk nuclei from near-defect sites. NV-hosting diamond provides an attractive platform to investigate these processes not only due to the promise of low-field, room-temperature DNP, but also because its comparatively simple composition makes it a clean platform to expose fundamental phenomena.

Capitalizing on these singular properties, this work explored the generation of $^{13}$C polarization under the combined action of optical and radio-frequency excitation. Unexpectedly, we observed inversion of the polarization signal under a strong RF drive far removed from the $^{13}$C Larmor frequency, a finding we interpreted as a form of 'solid effect' in an effective four-level subset of a three-spin $^{13}$C–NV–P1 model system. Unlike the standard case, however, here spin initialization derives from NV optical pumping, not temperature. More importantly, the reduced energy separations between different multi-electronic configurations allows one to drive the system via radio-frequency, not microwave (MW). The latter is made possible by nuclear spins, whose hyperfine interactions with defects mix the character of the system eigenfunctions to enable otherwise forbidden magnetic dipole transitions. Naturally, similar ideas also apply to more complex spin sets not considered here involving, e.g., more than two electron spins, though a quantitative description becomes increasingly involved. Since contributions from these ill-defined spin sets become dominant in a diamond crystal with a high concentration of nitrogen impurities such as ours, additional work in more suitable samples — including strong RF driving at other frequencies and RF-absorption spectra at varying magnetic fields — will be required to gain a fully quantitative description. Alternatively, optically detected magnetic resonance experiments on individual NV centers whose nuclear and electronic neighbors have been previously characterized[37-40] could provide a more controlled route to exposing the action of continuous RF excitation (even though the impact of this polarization on bulk spins becomes unobservable).

The ability to access the electronic bath through 'intra-band' transitions — i.e., transitions between electronic levels within a Zeeman-defined manifold — could perhaps be exploited to implement microwave-free routes to DNP, even in the absence of optical spin pumping. For example, assuming thermal initialization in the lowest electronic Zeeman manifold, different spin dynamics follow from an NV or a P1 spin flip, each taking place with a different unit time probability. Given a characteristic spin transport time away from the source electron/nuclear set, the above differences should lead to a net injection of nuclear spin magnetization into the bulk. A full analysis of this MW-free DNP process — involving not only the generation but also the transport of spin polarization — is presently under study and will be the subject of future work.

## ACKNOWLEDGMENTS

D.P. and C.A.M. acknowledge support from the National Science Foundation through grant NSF-1903839; they also acknowledge access to the facilities and research infrastructure of the NSF CREST IDEALS, grant number NSF-HRD-1547830. PRZ and RHA acknowledge financial support from CONICET (PIP-111122013010074 6CO), SeCyT-UNC (33620180100154CB) and ANPCyT (PICT-2014-1295).

## APPENDIX A: Model of synchronic inversion during spin diffusion

Here we consider (and rule out) a hypothetical scenario where polarization inversion emerges from partly synchronous RF excitation *during* spin diffusion from strongly hyperfine-coupled to bulk nuclei. Imagining a spectral chain where polarization flows from more to less strongly hyperfine-coupled sites (Fig. 5a), continuous RF inverts the observed NMR signal as polarization comes in and out of resonance with the applied RF; in other words, a sign reversal is a priori conceivable if the inverse polarization flow rate (i.e., the equivalent of a 'residence time' in the spectral chain) coincides with half the RF Rabi period.

Since the flow of magnetization is an intrinsically complex, many-body process, a quantum mechanical description is difficult. Instead, we build on the classical model introduced in Ref. [26] and reproduced in Fig. 5a. In this model, $m$ 'spectral boxes' represent groups of spins $\{N_i\}$, $i: 1 \dots m$ with given hyperfine couplings. Every spin in the $j$-th box has a resonance frequency lying within a range $\Delta \nu_j$ centered at the effective hyperfine shift $\|A_j\|$. Boxes are arranged following a decreasing order in the hyperfine shifts: The first box corresponds to strongly hyperfine-coupled nuclear spins, while the last box, $m$, corresponds to 'bulk' nuclear spins, i.e. those that are experimentally accessible.

We assume that only the first box is initially magnetized, i.e., 'populated'. Such a population has a positive 'character', which stands for a positive magnetization. The flow of magnetization occurs due to an inter-box transfer at rate $\gamma_{i,i+1}$. This flow is impacted by the RF irradiation, which can drive the magnetization of nuclear spins within some effective bandwidth $\delta \nu_b$. In terms of our model, the RF changes the character of the population in the irradiated box. This driving takes place at a flipping rate corresponding to the Rabi frequency, $\Omega_{\text{RF}}$. The action of the RF at a particular box does not affect the overall time scale at which the $m$-th box is populated. This means that positive and negative magnetization diffuse at the same rate. Instead, RF ultimately modulates the amount and sign of magnetization that reaches

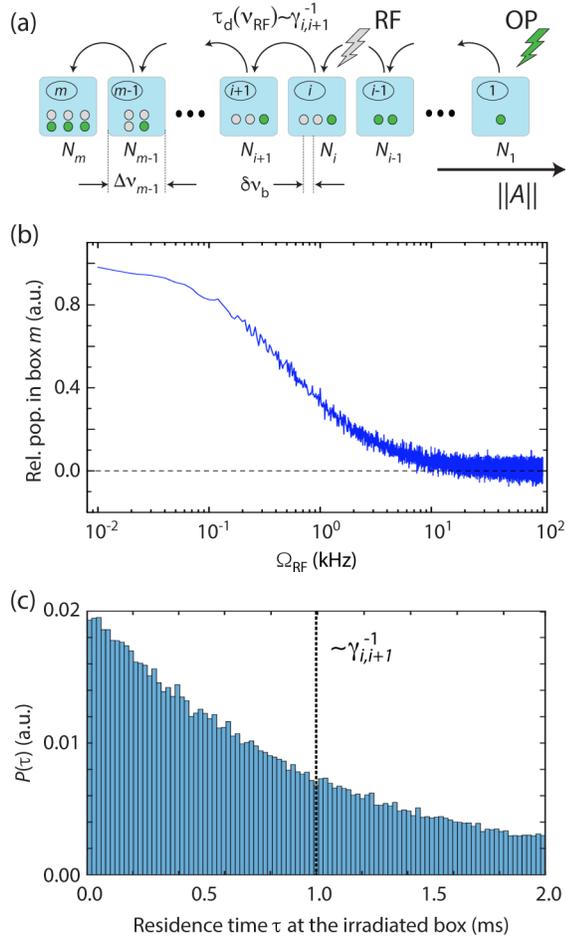

**Figure 5. Classical picture of spin diffusion.** (**a**) Spectral chain introduced in Ref. [26] to model magnetization transport from strongly-hyperfine coupled to bulk nuclear spins (see text). (**b**) Normalized population in the last box of the chain (corresponding to bulk magnetization) as a function of the RF Rabi amplitude $\Omega_{RF}$. (**c**) Normalized histogram of residence times at the irradiated box $i_0$. Vertical dashed line corresponds to the mean value of the distribution, which coincides with the inverse of the inter-box hopping rate $\gamma_{i,i+1}$. In (**b-c**) we used a 20-box model, with $\gamma_{i,i+1} = \gamma_0 = 1$ kHz $\forall i$, $i_0 = 8$ and $10^5$ Monte Carlo realizations for trajectory sampling.

the final box.

The time-evolution of the system can be evaluated by solving the appropriate set of differential equations or by Monte Carlo trajectory sampling. Using this last approach, we show in Fig. 5b the asymptotic population state of the $m$-th box as a function of $\Omega_{RF}$. For the sake of simplicity, we assume a homogeneous transfer rate, $\gamma_{i,i+1} = \gamma_0 = 1$ kHz $\forall i$. In addition, we consider that the RF irradiation acts on an intermediate box $i_0$ with $1 < i_0 < m$. We observe that for very weak driving $\Omega_{RF} \ll \gamma_0$, the last box is positively magnetized, since the time needed for flipping the character of the population at the irradiated box $i_0$ is much longer than the typical residence time at such box. In the opposite limit, $\Omega_{RF} \gg \gamma_0$, the character of the population at the irradiated box flips many times before it continues its flow towards the final box. Due to the intrinsically diffusive nature of magnetization transport, the flipping operation cannot occur in synchrony with the passage across the box $i_0$. Loosely speaking, a perfect $\pi$-pulse cannot take place at any intermediate stage of the magnetization transport.

Figure 5c reinforces the physical picture just described by showing the distribution of residence times at the irradiated box $i_0$, i.e. the amount of time the magnetization stays at such spectral box. Similar distributions are found for any other site along the spectral chain. Even though the mean value of such distribution precisely corresponds to $\gamma_0^{-1}$, it is clear that the flipping mechanism would sample the whole distribution of residence times, producing both positive and negative outcomes. So, either the final box is positively populated or its character averages to zero, which means an equal contribution of positive and negative population characters.

## APPENDIX B: Quantum model of spin dynamics under RF excitation

### B1. The spin Hamiltonian

We formalize here the quantum-mechanical few-level system description of the RF-induced signal reversal. We stress that this approach focusses exclusively on the *generation*, not the transport, of magnetization.

We start by considering a three-spin system, including a $^{13}$C nuclear spin **I** ($I = 1/2$), a NV center spin **S** ($S = 1$) and an electronic P1 spin **S'** ($S' = 1/2$). The complete Hamiltonian $H_T$ in the laboratory-frame is given by:

$$H_T = -\omega_0 I^z + \omega_e S^z + \omega_e S'^z + D(S^z)^2 + A_{zz} S^z I^z + A_{zx} S^z I^x + \mathcal{J}_d(S^+ S'^+ + S^- S'^-) \quad (B.1)$$

Here, $\omega_0 = \gamma_I B$, and $\omega_e = |\gamma_e| B$, while $D$ indicates the NV zero-field splitting. Coefficients $A_{zz}$ and $A_{zx}$ denote the (secularized) hyperfine tensor components coupling the $^{13}$C and NV spins, and $\mathcal{J}_d$ stands for the dipolar coupling strength between the NV and P1. Additionally, we assume the magnetic field direction is aligned with the NV axis and its modulus is tuned such that $2\omega_e \approx D$. This implies that the spin state $|0\uparrow\rangle$ for the NV-P1 pair is almost degenerate with $|-1\downarrow\rangle$. The degeneracy condition justifies the double-quantum terms in the NV-P1 'secular' dipolar interaction. Following the notation $|m_S, m_{S'}, m_I\rangle$ for the three-spin states, the explicit matrix representation of $H_T$ in the subspace spanned by the states $\{|0, +1/2, \uparrow\rangle, |0, +1/2, \downarrow\rangle, |-1, -1/2, \uparrow\rangle, |-1, -1/2, \downarrow\rangle\}$ is:

$$H_T = \begin{pmatrix} \frac{-\omega_0 + \omega_e}{2} & 0 & \mathcal{J}_d & 0 \\ 0 & \frac{\omega_0 + \omega_e}{2} & 0 & \mathcal{J}_d \\ \mathcal{J}_d & 0 & -\frac{(\omega_0 + A_{zz})}{2} + D - \frac{3\omega_e}{2} & -\frac{A_{zx}}{2} \\ 0 & \mathcal{J}_d & -\frac{A_{zx}}{2} & \frac{\omega_0 + A_{zz}}{2} + D - \frac{3\omega_e}{2} \end{pmatrix}$$
(B.2)

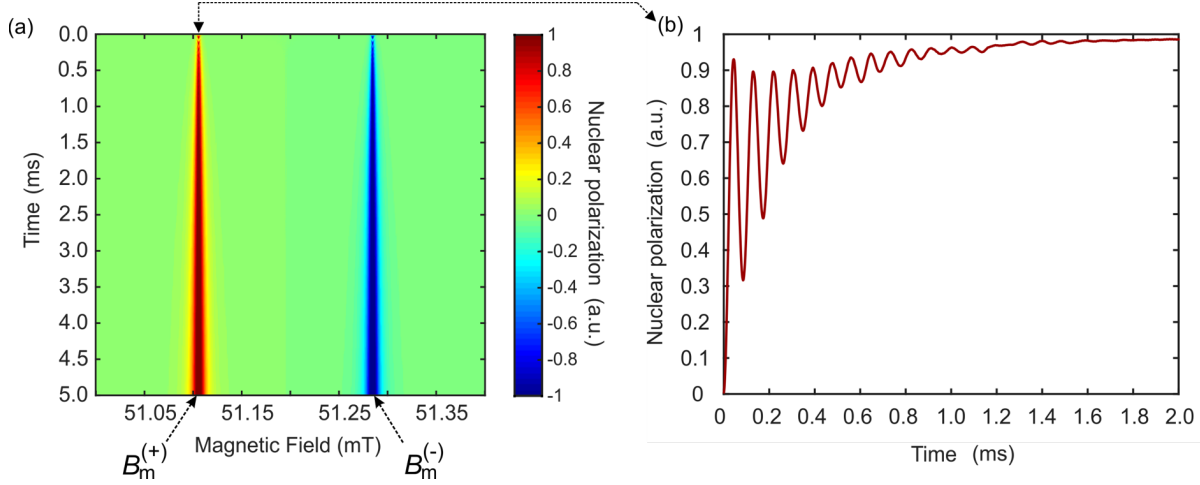

**Figure 6. Time dependence of nuclear polarization.** (**a**) $^{13}$C Nuclear spin polarization as function of time and the external magnetic field. Here, $A_{zz} = A_{zx} = 6.5$ MHz, $\mathcal{J}_d = 100$ kHz and the NV repolarization rate is 10 kHz. Each time-trace is averaged over $10^3$ Monte Carlo trajectories to account for the NV repolarization (see text). (**b**) Cross section of the dynamics shown in (**a**), i.e. single time-trace, at the particular field $B_m^{(+)} \approx 51.106$ mT.

Since we assume large hyperfine strengths, the natural quantization axis for nuclear spins in the $m_S = -1$ subspace has to be given by the hyperfine vector instead of the external magnetic field. This is not the case for $m_S = 0$, where the nuclear quantization axis is still given by the Zeeman term. Thus, we rotate the nuclear quantization axis to coincide with the vector

$$\vec{Z} = m_S A_{zx} \boldsymbol{i} + (m_S A_{zz} - \omega_0) \boldsymbol{k} \quad (B.3)$$

with the corresponding angle,

$$\tan \theta_{m_S} = \frac{m_S A_{zx}}{(m_S A_{zz} - \omega_0)} \quad (B.4)$$

and

$$\Delta_{m_S} = |\vec{Z}(m_S)| = \sqrt{(m_S A_{zx})^2 + (m_S A_{zz} - \omega_0)^2}. \quad (B.5)$$

The matrix representation of the Hamiltonian after this rotation is given by

$$H_T' = \begin{pmatrix} \frac{-\omega_0 + \omega_e}{2} & 0 & \mathcal{J}_d c_1 & -\mathcal{J}_d s_1 \\ 0 & \frac{\omega_0 + \omega_e}{2} & \mathcal{J}_d s_1 & \mathcal{J}_d c_1 \\ \mathcal{J}_d c_1 & \mathcal{J}_d s_1 & -\frac{\Delta_{-1}}{2} + D - \frac{3\omega_e}{2} & 0 \\ -\mathcal{J}_d s_1 & \mathcal{J}_d c_1 & 0 & \frac{\Delta_{-1}}{2} + D - \frac{3\omega_e}{2} \end{pmatrix},$$

(B.6)

where $c_1 = \cos(\theta_{-1}/2)$ and $s_1 = \sin(\theta_{-1}/2)$. Notice that the NV-P1 dipolar coupling term leads to relatively small, off-diagonal matrix elements (of order $\sim \mathcal{J}_d$). In what follows, the nuclear states in the subspace $m_S = -1$ are denoted with primes, $\uparrow'$ and $\downarrow'$, as a reminder that they are referenced to $\vec{Z}$.

The effect of optical pumping is introduced here by means of a projective, non-unitary operation. With that purpose, we enlarge the Hilbert space and introduce two ancillary states, $|0, -1/2, \uparrow\rangle$ and $|0, -1/2, \downarrow\rangle$. Then, we implement a Monte Carlo quantum jump approach that stochastically projects

$$|-1, -1/2, \uparrow'\rangle \to |0, -1/2, \uparrow\rangle$$

and

$$|-1, -1/2, \downarrow'\rangle \to |0, -1/2, \downarrow\rangle.$$

The probability of such a projection corresponds to the NV repolarization rate per unit time, and each trajectory is averaged over a sufficiently large number of realizations.

Using the above framework, in Fig. 6a we evaluate the time dependence of the $^{13}$C nuclear polarization as a function of time and the external magnetic field. As in Fig. 1b, we identify the two matching fields $B_m^{(+)}$ and $B_m^{(-)}$ for the simplified three-spin system defined above. In particular, we show in Fig. 6b the build-up of nuclear polarization at $B_m^{(+)}$, which corresponds to the solution of the nonlinear equation

$$\frac{\omega_0 + \omega_e}{2} = -\frac{\Delta_{-1}}{2} + D - \frac{3\omega_e}{2}. \quad (B.7)$$

This equation represents the quasi-degeneracy condition between states $|0, 1/2, \downarrow\rangle$ and $|-1, -1/2, \uparrow'\rangle$, and its precise solution ultimately depends on the particular hyperfine coupling.

**B2. RF irradiation**

Now we want to consider the effect of a time-dependent term that describes the effect of RF excitation,

$$H_{RF} = -\gamma_C B_1 I^x \cos(\omega t) = -\Omega_{RF} I^x \cos(\omega t) \quad (B.8)$$

where the frequency $\omega$ is set to drive strongly-hyperfine shifted nuclear spins (i.e., $\omega \gg \omega_0$). In order to derive an effective, time-independent Hamiltonian we stay at a fixed magnetic field, which corresponds to $B_m^{(+)}$ for a particular hyperfine coupling. For (large) hyperfine couplings ($\gg \omega_0$) where the degeneracy condition (Eq. B.8) does not hold, a detuning $\delta$ can be associated to the energy difference between states $|0,1/2,\downarrow\rangle$ and $|-1,-1/2,\uparrow'\rangle$. For simplicity, we assume $\delta \ll \Delta_{-1}$. We stress, however, that the validity of our observations does not rely on this assumption.

The above considerations lead to a well-defined hierarchy of states. The energies of the first three states, $|0,1/2,\uparrow\rangle$, $|0,1/2,\downarrow\rangle$ and $|-1,-1/2,\uparrow'\rangle$, are assumed very close to each other. In fact, these energies differ in $\sim max\{\omega_0, \delta\}$. The remaining state, $|-1,-1/2,\downarrow'\rangle$, is energetically distant by an amount equal to $\Delta_{-1}$. Since, in general, $\omega \sim \Delta_{-1}$, the natural unitary transformation that defines the rotating frame in this system is given by

$$R = \exp\{-i\mathbb{W}t\}, \quad (B.9)$$

where

$$\mathbb{W} = \begin{pmatrix} 0 & 0 & 0 & 0 \\ 0 & 0 & 0 & 0 \\ 0 & 0 & 0 & 0 \\ 0 & 0 & 0 & \omega \end{pmatrix}. \quad (B.10)$$

Before computing the lowest order in Average Hamiltonian Theory[41,42] (AHT), we diagonalize $H_T'$ by means of the appropriate rotation $\mathcal{U}$ into its eigen-frame. This rotation retains the hybridization of states induced by the off-diagonal matrix elements $\propto \mathcal{J}_d$, and thus avoids losing it in the lowest-order AHT averaging. The set of exact eigenstates is labeled by $\{|a\rangle, |b\rangle, |c\rangle, |d\rangle\}$, and since the mixing is weak, we can safely associate

$$\begin{aligned} |a\rangle &\dashrightarrow |0,+1/2,\uparrow\rangle \\ |b\rangle &\dashrightarrow |0,+1/2,\downarrow\rangle \\ |c\rangle &\dashrightarrow |-1,-1/2,\uparrow'\rangle \\ |d\rangle &\dashrightarrow |-1,-1/2,\downarrow'\rangle. \end{aligned} \quad (B.11)$$

Finally, following the standard AHT recipe,

$$\mathcal{H} = R[\mathcal{U}(H_T' + H_{RF})\mathcal{U}^{-1}]R^{-1} - iR\dot{R}^{-1} \quad (B.12)$$

and, since $\mathcal{U}H_T'\mathcal{U}^{-1}$ is diagonal by definition, we obtain

$$\mathcal{H} = \mathcal{U}H_T'\mathcal{U}^{-1} + R\mathcal{U}H_{RF}\mathcal{U}^{-1}R^{-1} - \mathbb{W} \quad (B.13)$$

The effective Hamiltonian we employed for simulating the effect of RF irradiation is given by the zeroth-order in AHT,

$$H_{\text{eff}}(\omega, \Omega_{RF}) = \frac{1}{T}\int_0^T \mathcal{H}(t')dt', \quad (B.14)$$

with $T = 2\pi/\omega$. Notice that the only time-dependent term in $\mathcal{H}(t')$ corresponds to $R\mathcal{U}H_{RF}\mathcal{U}^{-1}R^{-1}$.